\RequirePackage{lineno}
\documentclass[aps,prb,preprint,superscriptaddress]{revtex4}

\usepackage[a4paper]{geometry}
\usepackage{graphicx}
\usepackage{bm}
\usepackage{amsmath}
\usepackage{amssymb}
\usepackage{hyperref}
\usepackage{xcolor}

\begin{document}

\title{An atomistic model for the charge distribution in layered MoS$_{2}$}
	
\author{Yida Yang}
\affiliation{Guangxi Key Laboratory for Relativistic Astrophysics, Department of Physics, Guangxi University, Nanning 530004, P. R. China.}

\author{Michel Devel}
\affiliation{FEMTO-ST institute, UBFC, CNRS, ENSMM, 15B avenue des Montboucons, 25030 Besan\c{c}on CEDEX, France}

\author{Zhao Wang}
\email{zw@gxu.edu.cn}
\affiliation{Guangxi Key Laboratory for Relativistic Astrophysics, Department of Physics, Guangxi University, Nanning 530004, P. R. China.}

\begin{abstract}
We present an atomistic model for predicting the distribution of doping electric charges in layered molybdenum disulfide (MoS$_{2}$). This model mimics the charge around each ion as a net Gaussian-spatially-distributed charge plus an induced dipole, and is able to predict the distribution of doping charges in layered MoS$_{2}$ in a self-consistent scheme. The profiles of doping charges in monolayer MoS$_{2}$ flakes computed by this charge-dipole model are in good agreement with those obtained by density-functional-theory calculations. Using this model, we quantitatively predict the charge enhancement effect in MoS$_{2}$ monolayer nanoribbons, with which strong ionic charge-localization effects are shown.
\end{abstract}
		
\maketitle


\section{Introduction}
Two-dimensional (2D) materials are ideal candidates for nanoelectromechanical systems (NEMS) thanks to their unique electronic, optical and mechanical properties and peculiar structures.\cite{Butler2013} 2D layered MoS$_{2}$ has recently been used as main components in various devices including sensors,\cite{Huang2015} actuators,\cite{Acerce2017} resonators,\cite{Lee2013} piezoelectric generators,\cite{Maity2017} supercapacitor,\cite{Acerce2015} and field-emission devices.\cite{Wu2009a} The knowledge of the distribution of electric charges in the layered MoS$_{2}$ is a key aspect for understanding the damage mechanism and stability criteria in device components during charging, and is hence critical for the design of electromechanical devices since doping charges could strongly influence the electromechanical coupling,\cite{Gartstein2003} electronic band structures,\cite{Jung2009} charge screening\cite{Guinea2007} or field emission\cite{Purcell2002} properties of the component material.

Experimentally, electrostatic force microscopy (EFM) and Kelvin force microscopy (KFM) have been used to image the charge distribution in nanostructures such as carbon nanotubes (CNTs)\cite{Brunel2010} and graphene.\cite{Datta2009} Electric charges in nanomaterials were found to accumulate at the edges due to strong Coulomb repulsion.\cite{Wang2008c,Wang2009} Density functional theory (DFT) calculations have been established for the theoretical interpretation of this effect,\cite{Keblinski2002} however not in the range of dimensions often accessible by experiments due to the breakdown of periodic symmetry. It is hence critical to develop a model at larger scale for accurately predicting the charge distribution in nanostructures of size comparable to those of the samples used in experiments. Moreover, it is highly desirable that this model could provide an atomistic description of the systems in order to combine with empirical force fields for describing coupled electrical and mechanical effects\cite{Brennan2017,Hartman2004,Wange2009,Wangd2009,Bennett2010} in finite-size nanostructures by atomistic simulations.\cite{Wang2007,Wang2007a}

Recently, a Gaussian-regularized atomistic model has been developed to study electrostatic effects in carbon nanomaterials based on the atomic dipole theory of Applequist \textit{et al.}\cite{Applequist1972} and the electrostatic polarization model of Jensen \textit{et al.} \cite{Jensen2002,jensen2004} and Mayer.\cite{Mayer2006} This charge-dipole (QP) model has recently been used to predict the charge distribution in CNTs and was validated by EFM experiments.\cite{Wang2008c} In the present work, we extend this model to layered MoS$_{2}$ taking the ionic electrostatic interactions between atoms of different types into account, thanks to parameters obtained through DFT calculations. This model provides an atomistic description for the self-consistent electrostatic interactions between the atomic charges, dipoles and external electric fields, and is capable of dealing with relatively-large systems.

The outline of this paper is as follows. Details about the DFT calculations and QP model are presented in Section II. A comparison to DFT calculation results is presented in Section III. Finally, the charge enhancement effect in MoS$_{2}$ monolayer is predicted in Section IV. We draw conclusions in Section V.

\section{Methods}

\subsection{Density Functional Theory calculations}

\begin{figure}[htp]
\centerline{\includegraphics[width=9cm]{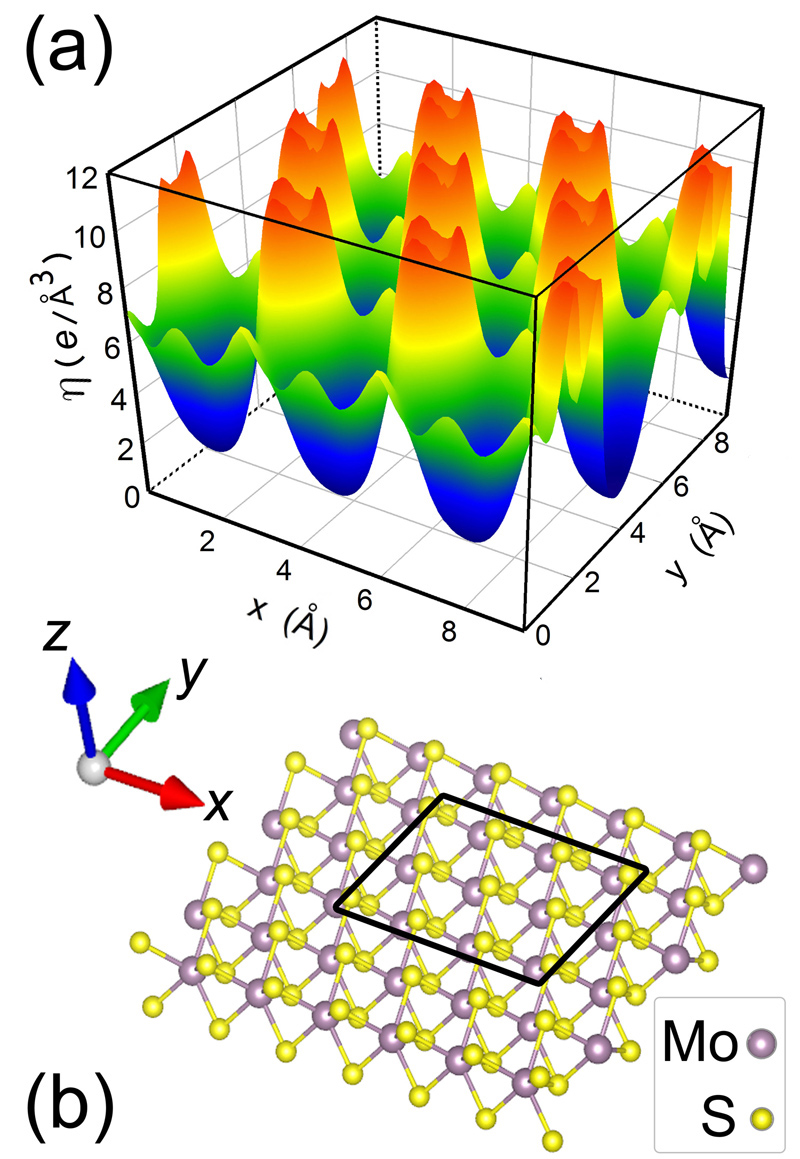}}
\caption{\label{F1}
(a)  Density profile of the intrinsic electric charge in an infinite pristine MoS$_{2}$ monolayer. (b) Atomistic structure of the monolayered MoS$_{2}$. The solid lines highlight the squared zone in which the charge density profile is depicted in (a).
}
\end{figure}

DFT calculations are conducted within the framework of spin-polarized plane-wave density functional theory (PW-DFT), as implemented in the Vienna \textit{ab-initio} simulation package (VASP).\cite{Kresse1996,Kresse1999} The generalized gradient approximation (GGA) with the Perdew-Burke-Ernzerhof (PBE) functional and projector augmented wave (PAW) pseudo-potentials are used. We adopt a $2 \times 2 \times 1$ supercell.
The vacuum size is set to be larger than $15$\;\AA $ $ between two adjacent images. An energy cutoff of $400\;\mathrm{eV}$ is used for the plane-wave expansion of the electronic wave function. The lattice structure is relaxed by the conjugated gradient algorithm. The 2D Brillouin zone integration using the $\Gamma$-center scheme is applied with a $6 \times 6$ grid for geometry optimization, and a $7 \times 7 \times 7$ grid for static electronic structure calculations in the Monkhorst-Pack scheme.

The density profile of the \textit{intrinsic} electric charges in an infinite pristine monolayer of MoS$_{2}$ is depicted in Fig.\ref{F1}. A strong ionic charge-localization effect can be observed. \textit{i.e.} the electric charge is found to accumulate on the sites of S ions forming a volcanic-cone-like profile. The concave at the sites of the S atom is caused by the repulsive interaction with valence electrons, while this is not observed on the charge profile of the Mo atoms. Note, that the density of the intrinsic electric charge is much higher than that of the doping charge shown in the figures below.

\begin{figure}[htp]
\centerline{\includegraphics[width=13cm]{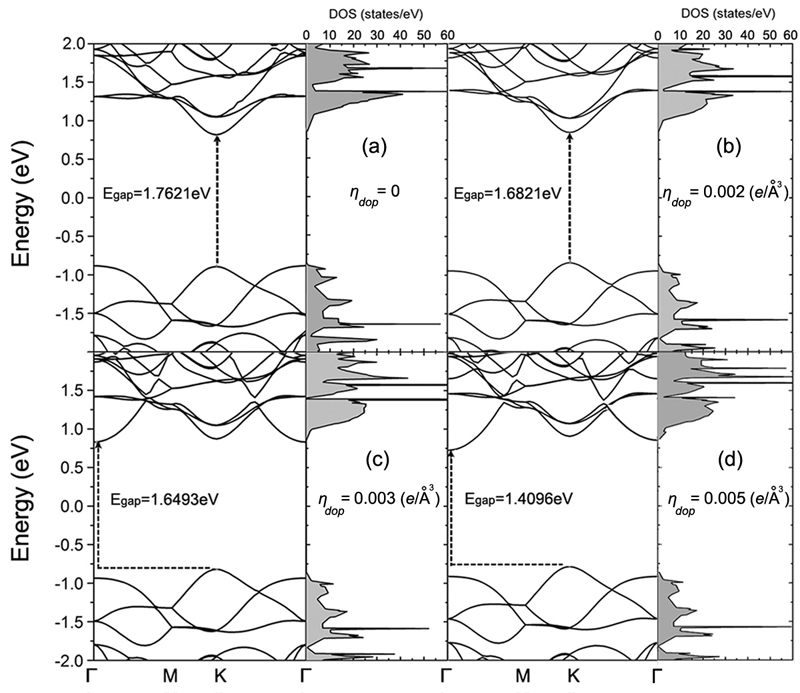}}
\caption{\label{F2}
Electronic band structure and density of state of an infinite monolayer MoS$_{2}$ that is doped with a charge density of 0 (a), 0.002 (b), 0.003 (c) and 0.005 (d) electron/\AA$^{3}$.
}
\end{figure}

The pristine monolayer MoS$_{2}$ is then subjected to a quantity of doping charge with a global density $\eta_{dop}$. Fig.\ref{F2} shows the electronic band structure (EBS) and density of state (DOS) at different doping levels. It can be seen that the EBS of MoS$_{2}$ starts to be significantly modified and direct-to-indirect band-gap switch can be observed when $\eta_{dop}$ goes beyond $0.002\;e/$\AA$^{3}$. The computation done below is thus controlled with $Q_{dop}<0.002\;e/$\AA$^{3}$ in order to avoid significant modification to EBS and DOS, which would increase uncertainty in the transferability of the subsequent parameterization of the charge-dipole model. Note that benchmarks were performed on an infinite pristine sample computing its DOS and band gap, and good agreement was obtained with data provided in the literature, as shown in supplementary material.

\subsection{Gaussian-regularized charge-dipole model}

In the charge-dipole (QP) model, each atom is associated with an electric charge $q$ and an induced dipole $\bm{p}$. The total electrostatic energy $U$ for a system composed of $N$ atoms can be written as follows,

\begin{multline}
\label{Eq1}
U^{elec} =\sum_{i=1}^N{q_i(\chi_i+V_i)}-\sum_{i=1}^N{\bm{p}_i\cdot\bm{E_i}}+
\frac{1}{2}\sum_{i=1}^N{\sum_{\substack{j=1}}^N{q_i T^{i,j}_{q-q} q_j}}\\-\sum_{i=1}^N{\sum_{\substack{j=1}}^N{\bm{p}_i \cdot \bm{T}^{i,j}_{p-q} q_j}}
-\frac{1}{2}\sum_{i=1}^N{\sum_{\substack{j=1}}^N{\bm{p}_i \cdot \bm{T}^{i,j}_{p-p} \cdot \bm{p}_j}}
\end{multline}

\noindent where $\chi_i$ is usually called electronegativity of the atom/ion $i$ (though it is rather an electronegativity divided by the proton charge, if the electronegativity is defined by the partial derivative of a given isolated atom/ion energy with respect to the electron number), $V_i$ and $\bm{E_i}$ stand for the external potential and electric field, respectively, at the location of atom/ion $i$. $T$ and $\bm{T}$ usually are the electrostatic interaction tensors between point charges or dipoles in vacuum, which allow to compute the electrostatic potential or field at a point $\bm{r_i}$ created by a point source (charge or dipole) located at $\bm{r_j}$. They are defined as $T^{i,j}_{q-q}=1 / 4 \pi \varepsilon_{0} r_{ij}$, $\bm{T}^{i,j}_{p-q}=-\nabla_{\bm{r}_i} T^{i,j}_{q-q}$ and $\bm{T}^{i,j}_{p-p}=-\nabla_{\bm{r}_j} \otimes \nabla_{\bm{r}_i} T^{i,j}_{q-q}$, where $r_{i,j}=\left| \bm{r}_{i}-\bm{r}_{j} \right|$. For point charges or point dipoles, the terms $i=j$ in the double-summations are usually respectively connected to the chemical hardness and polarizability of the corresponding atoms. However, in the present model, the charges and dipoles are not considered to be point-like but to correspond to spherically symmetric, radially Gaussian, electronic charge distributions. This avoids divergence problems such as ``polarization catastrophes'' due to the fact that in covalent bonds the electronic clouds are overlapping, by replacing the standard vacuum $T$ and $\bm{T}$ defined above by their convolution with two Gaussian distributions of the type $\exp(-r^2/R^2)/\pi^{3/2}R^3$,\cite{Jensen2002,Langlet2006,Mayer2006}

\begin{equation}
\label{Eq2}
\begin{array}{c}
\left\{
\begin{array}{lcl}
T^{i,j}_{q-q} & = &\frac{1}{4\pi\epsilon_0 r_{i,j}} \rm{erf}\left(\frac{r_{i,j}}{\sqrt{R_i^2+R_j^2}}\right)\\
\bm{T}^{i,j}_{p-q}&=&-\bm\nabla_{\bm{r}_i} T^{i,j}_{q-q}=\frac{1}{4\pi\epsilon_0}\frac{\bm{r}_{i,j}}{r_{i,j}^3}\left[\rm{erf}\left(\frac{r_{i,j}}{\sqrt{R_i^2+R_j^2}}\right)-\frac{2}{\sqrt\pi}\frac{r_{i,j}}{\sqrt{R_i^2+R_j^2}}
 \exp\left(-\frac{r_{i,j}^2}{R_i^2+R_j^2}\right)\right]\\
\bm{T}^{i,j}_{p-p}&=&-\bm\nabla_{\bm{r}_j} \otimes \bm\nabla_{\bm{r}_i} T^{i,j}_{q-q} \\
 &=& \frac{1}{4\pi\epsilon_0}\left\{\frac{3\bm{r}_{i,j} \otimes \bm{r}_{i,j}-r_{i,j}^2\bm{I}}{r_{i,j}^5}\left[\rm{erf}\left(\frac{r_{i,j}}{\sqrt{R_i^2+R_j^2}}\right)-\frac{2}{\sqrt\pi}\frac{r_{i,j}}{\sqrt{R_i^2+R_j^2}}
 \exp\left(-\frac{r_{i,j}^2}{R_i^2+R_j^2}\right)\right]\right. \\
& &\left. -\frac{4}{\sqrt{\pi}}\frac{\bm {r_{i,j}}\otimes \bm{r_{i,j}}}{r_{i,j}^2}\frac{1}{({\sqrt{R_i^2+R_j^2}})^3}\exp\left(-\frac{r_{i}^2}{R_i^2+R_j^2}\right)\right\}%
\end{array}.\right.\\
\forall{i \neq j}
\end{array}
\end{equation}

\noindent where $\bm{r}_{ij}= \bm{r}_{i}-\bm{r}_{j}$ is the vector pointing from ion $j$ to $i$, and $R_i$ and $R_j$ are the width of the Gaussians charge distributions for ions $i$ and $j$ respectively, which would vary with the type and position of the ions. This allows to remove divergences (when $i=j$, i.e. $\lim{r_{i,j}\rightarrow 0}$) and express self-terms as:

\begin{equation}
\label{Eq3}
\begin{array}{c}
\left\{
\begin{array}{ll}
q_{i}T^{i,i}_{q-q}q_{i} =
\frac{q^2_i}{4\pi\epsilon_0}\frac{\sqrt{2/\pi}}{R_i}\\
\bm{p}_{i}\cdot\bm{T}^{i,i}_{p-q}q_{i} = 0\\
\bm{p}_{i}\cdot\bm{T}^{i,i}_{p-p}\cdot\bm{p}_{i} = -\frac{ p^2_i}{4\pi\epsilon_0}\frac{\sqrt{2/\pi}}{3R_i^3}.
\end{array}\right.\\
\end{array}
\end{equation}

Periodic boundary conditions (PBC) can be included in this model by adding periodic images to the propagators (Eq.\ref{Eq2}) taking $r_{ij} = r_i -  r_j + k*a$, where $a$ is the periodic length in a given direction, $k = -m,-m+1,-m+2,...,-1,0,1,...,m-2,m-1,m$ with $m$ being a large integer. PBC were used for our computations on infinite nanoribbons but not on flakes. Note that a generalization of the charge-dipole model to systems with different atoms has been provided in Ref.32. Moreover, charge equilibration models are known to result in unreasonable charge distributions predicted for geometries far from equilibrium due to incorrect description to long-range charge transfer. \cite{Chen2007,Chen2008,Bredas2004} This problem persists even for time-dependent density functional theory.\cite{Kummel2017} Note that all geometries used in the present work are relaxed to be in full-equilibrium to avoid such a problem.

Since the equilibrium charges and dipoles should correspond to the global minima of $U^{elec}$, its derivatives with respect to the $q_i$ and $\bm{p_i}$ should therefore be zero. Furthermore, the conservation of the total molecular net charge $Q^{tot}$ can be imposed self-consistently by using a Lagrange multiplier $\lambda$ and minimizing $U^{elec}-\lambda (\sum_{j=1}^N{q_{j}}-Q^{tot})$.\cite{Mayer2007} We note that multiple $\lambda$ can be involved if charge conservation must be enforced for a system composed of several separated molecules and that $\lambda$ can also be interpreted as an ``instantaneous electronegativity'' common to all atoms at electric equilibrium.\cite{Ma2006} These boundary conditions enable us to obtain the equilibrium configurations of the charges and dipoles by solving $N$ linear vectorial equations and $N+1$ linear scalar equations (corresponding to a square matrix of order $4N+1$).

\begin{equation}
\label{Eq4}
\begin{array}{c}
\left\{ \begin{array}{ll}
\sum\limits_{j=1}^N{\bm{T}^{i,j}_{p-p} \bm{p}_{j}}+\sum\limits_{j=1}^N{\bm{T}^{i,j}_{p-q}q_{j}}=-\bm{E_i}\\
\sum\limits_{j=1}^N{\bm{T}^{i,j}_{p-q}\cdot\bm{p}_{j}}+\sum\limits_{j=1}^N{T^{i,j}_{q-q}q_{j}}-\lambda = -(\chi_i+V_i)\\
\sum\limits_{j=1}^N{q_{j}}=Q^{tot}
\end{array}\right.\\
\forall{i=1,...,N}
\end{array}
\end{equation}

\noindent Key parameters including the Gaussian charge distribution width $R_i$ and electronegativity $\chi_i$ are obtained respectively for Mo and S atoms by fitting to the charge distributions obtained from DFT calculations, as detailed below.

\begin{figure}[htp]
\centerline{\includegraphics[width=12cm]{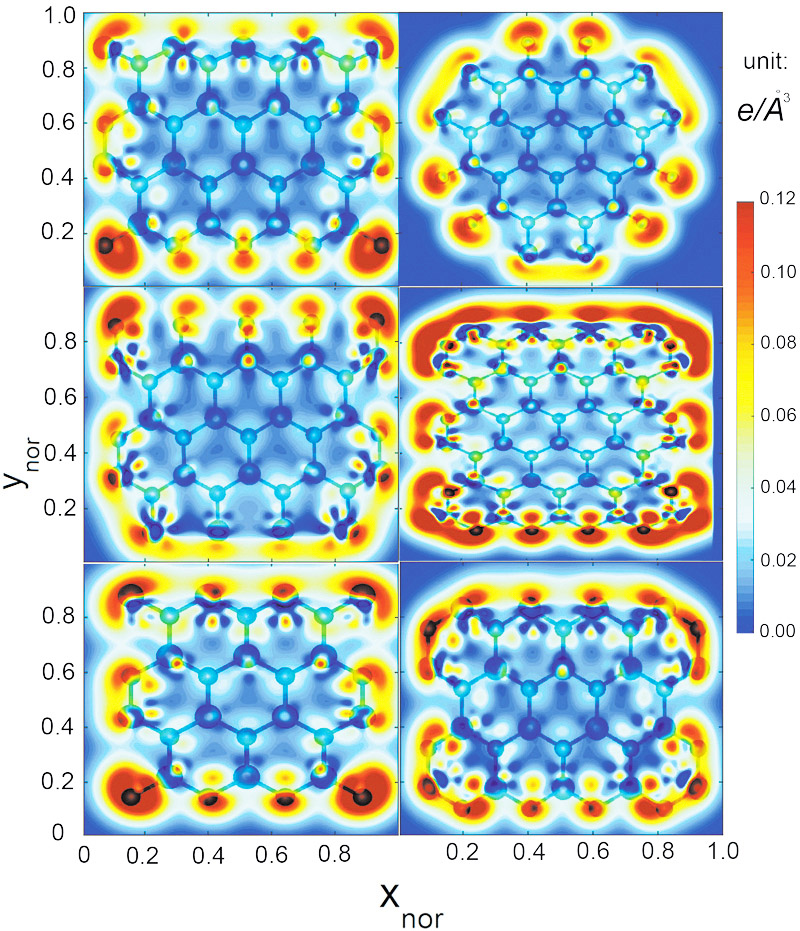}}
\caption{\label{F3}
DFT-calculated density profiles of doping charge in different MoS$_{2}$ monolayer sheets that are doped with an electron. The watermarked circles represent the in-plane positions of corresponding S and Mo atoms. $x_{nor}$ and $y_{nor}$ stand for the in-plane coordinates that are normalized with respect to the sheet width and length.
}
\end{figure}

\begin{table}[h]
\caption{\label{table1} Values of the Gaussian charge density widths and atomic electronegativities.}
\begin{center}
\begin{tabular}{c|c|c}
& R (\AA )& $\chi$ (V)            \\
\,\,atom\,\,&\,\,inner\,\,\,\,\,\,edged\,\,&\,\,inner\,\,\,\,\,\,edged\,\,\\
\hline
\,\,S\,\,&\,\,0.2118\,\,\,\,\,\,0.2616\,\,&\,\,2.0267\,\,\,\,\,\,1.6657\,\,\\
\,\,Mo\,\,&\,\,0.7019\,\,\,\,\,\,0.8626\,\,&\,\,-1.0948\,\,\,\,\,\,-1.7686\,\,
\end{tabular}
\end{center}
\end{table}

Two different sets of DFT calculations are conducted to compute the distributions of intrinsic and doping electric charges in monolayered MoS2 flakes as shown in Fig.\ref{F1} and Fig.\ref{F3} for example, respectively. The results are used to estimate the values of the Gaussian charge distribution widths $R_i$ in the QP model (Eq.\ref{Eq2}), with an analytical expression of the electronegativity $\chi_i$ which uses the atomic charges computed by a Bader-type analysis, \cite{Henkelman2006}

\begin{equation}
\label{Eq5}
\chi_i=\sum\limits_{j=1}^N{T^{i,j}_{q-q}q_{j}}
\end{equation}

\noindent by which the values of $\chi_i$ do not need to be estimated before the determination of $R_i$. These $R_i$ are first roughly estimated by fitting Gaussian functions to the DFT-calculated average radial atomic charge densities. Then, an iterative-correction algorithm is used to determine the exact best-fitting value of $R_i$ by numerically fitting all the atomic total charge density profiles calculated by the QP model to those computed by DFT, as shown in the Supplementary Materials. For each type of ions, two different values of $R_i$ are obtained as follows. One for bulk-positioned ions that are characterized by the same number of nearest neighbors as for an atom in an infinite MoS$_{2}$ monolayer (\textit{inner} denoted), and another for those with a reduced number of nearest neighbors due to edge positions (\textit{edge} denoted). The obtained values of $R_i$ are listed in Table \ref{table1} for S and Mo atoms, respectively. We see that $R_i$ is larger for the edged atom, this is similar to the Gaussian charge distribution widths in $sp^{2}$-hybridized carbon nanomaterials.\cite{Mayer2007} It is also found that the $R_i$ values of Mo are larger than those of S anions.

To determine the values of $\chi_i$, we input DFT-calculated intrinsic charge distribution into Eq.\ref{Eq5}. $\chi_i$ is a complex function that varies with the size of the MoS$_{2}$ monolayer and the environment of a given atom/ion, but $\chi_i$ converges at large size. For the model simplicity, the convergent values of the electronegativities for each kind of atom, in relatively large layers, are therefore used as parameters for the QP model and listed in Table \ref{table1}. We see that $\chi$ of Mo in layered MoS$_{2}$ is comparable to that of the bulk $-2.16$, while that of S is below the bulk value of $2.58$. Note, that the intrinsic dipoles are neglected in the estimation of $\chi_i$ due to the difficulty in determining the intrinsic dipole from DFT-calculated 3D charge distribution. This would hold as an approximation since the contribution of intrinsic dipoles to intrinsic fields is usually minor compared to that of net charges. However, it seems probable that the values of the calculated QP dipoles effectively compensates for the approximations in the determination of the parameters which is based solely on charges.

Further details about the computation of $R$ and $\chi$ parameters are provided in the supplementary material.

\section{Comparison to DFT}

\begin{figure}[htp]
\centerline{\includegraphics[width=12cm]{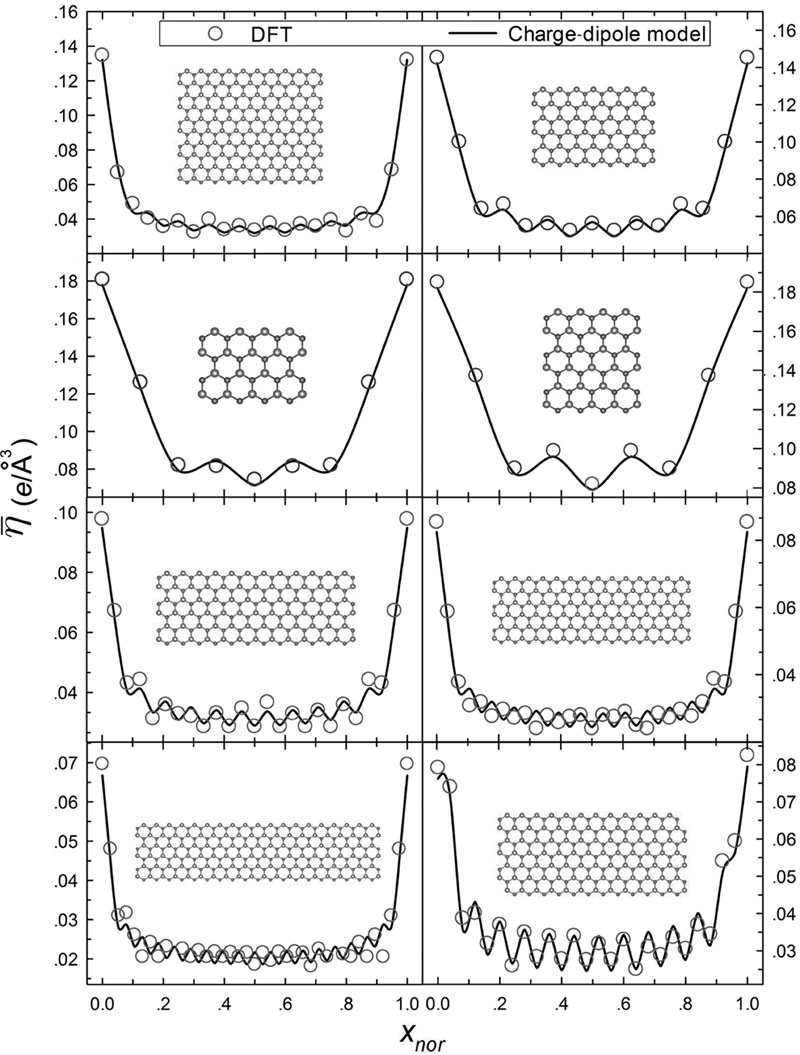}}
\caption{\label{F4}
Average density profile $\bar{\eta}$ of a doping electric charge in MoS$_{2}$ monolayer flakes along the longitudinal axis $x$. Comparison between results obtained by DFT calculations (symbols) and the charge-dipole model (lines). The $x$ positions (abscissa axis) are normalized with respect to the sheet length.}
\end{figure}

\begin{figure}[htp]
\centerline{\includegraphics[width=12cm]{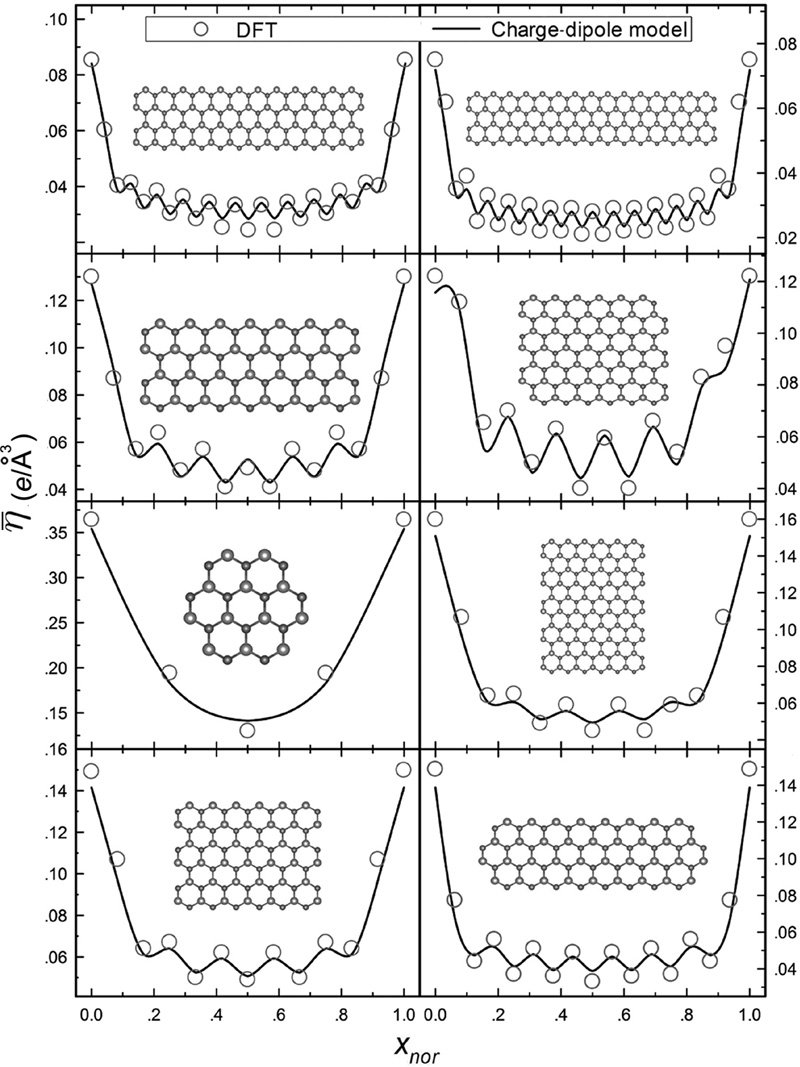}}
\caption{\label{F4b}
Average density profile $\bar{\eta}$ in doped MoS$_{2}$ monolayer flakes along $x$ axis. Comparison between results obtained by DFT calculations (symbols) and the charge-dipole model (lines). The $x$ positions (abscissa axis) are normalized with respect to the sheet length.}
\end{figure}

A comparison is made between the distributions of a doping electron computed by the QP model and another set of DFT calculations on relatively large MoS$_{2}$ flakes, as shown in Fig.\ref{F4} and \ref{F4b}. We see that the agreement on the average charge density of the doping charge $\eta_{dop}$ is remarkable, signifying that the redistribution of the doping charge in MoS$_{2}$ monolayer can be well captured by the QP model. It is shown that the density of doping charge is enhanced at the flake edge, similar to that predicted for CNTs.\cite{zw2010d} However, unlike in CNTs, the charge profile in MoS$_{2}$ oscillates due to the aforementioned ionic charge-localization effects. This is an unique electrostatic feature of ionized nano-crystals.

\section{Predictions of charge enhancement}
\begin{figure}[htp]
\centerline{\includegraphics[width=10cm]{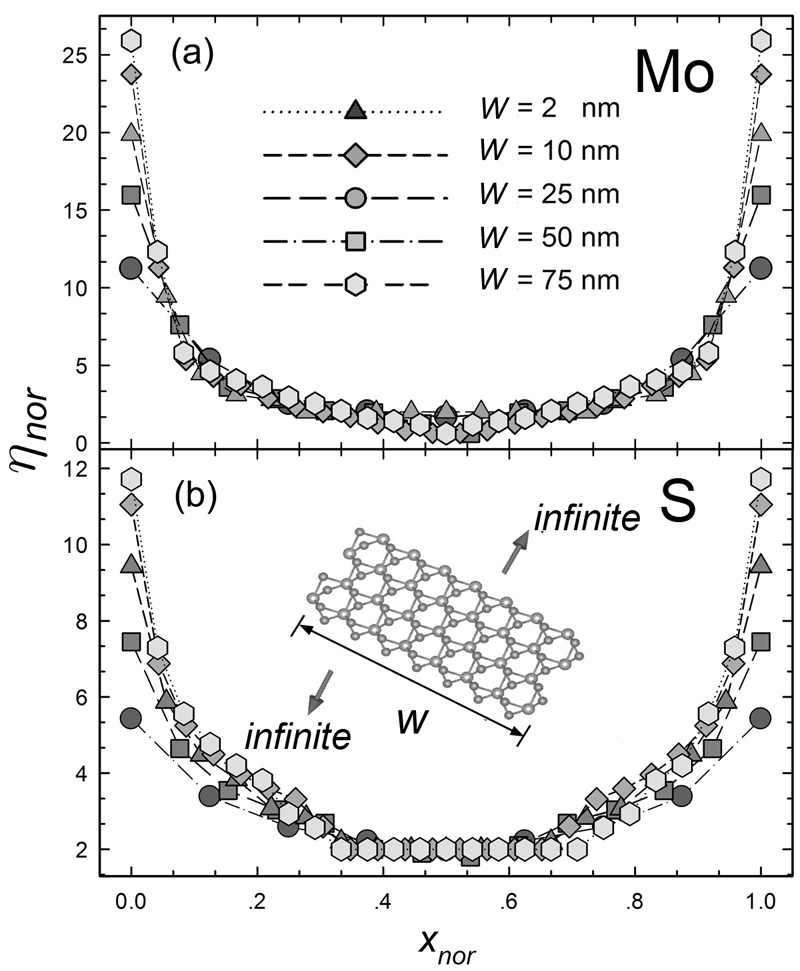}}
\caption{\label{F5}
Profile of the normalized density $\eta_{nor}$ of a doping electric charge in monolayer MoS$_{2}$ nanoribbons of an infinite length and a finite width $W$. $\eta_{nor}$ is normalized with respect to $\eta$ at the ribbon center. The abscissa axis is normalized with respect to $W$.}
\end{figure}

We use the parametrized QP model to quantitatively predict the charge enhancement effect in monolayer MoS$_{2}$ nanoribbons, which is a significant feature of two-dimensional materials for energy storage\cite{Acerce2015} and field-emission applications.\cite{Purcell2002} To generalize our results for the size of samples commonly used in experiments, it is interesting to investigate infinitely-long sheets or strips. We therefore compute the distribution of net electric charges in MoS$_{2}$ nanoribbons infinite in length of different widths $W$, as shown in Fig.\ref{F5}. We see that the charge enhancement at the edges is more significant for longer sheets. This behavior is comparable to that in CNTs.\cite{Keblinski2002}

\begin{figure}[htp]
\centerline{\includegraphics[width=10cm]{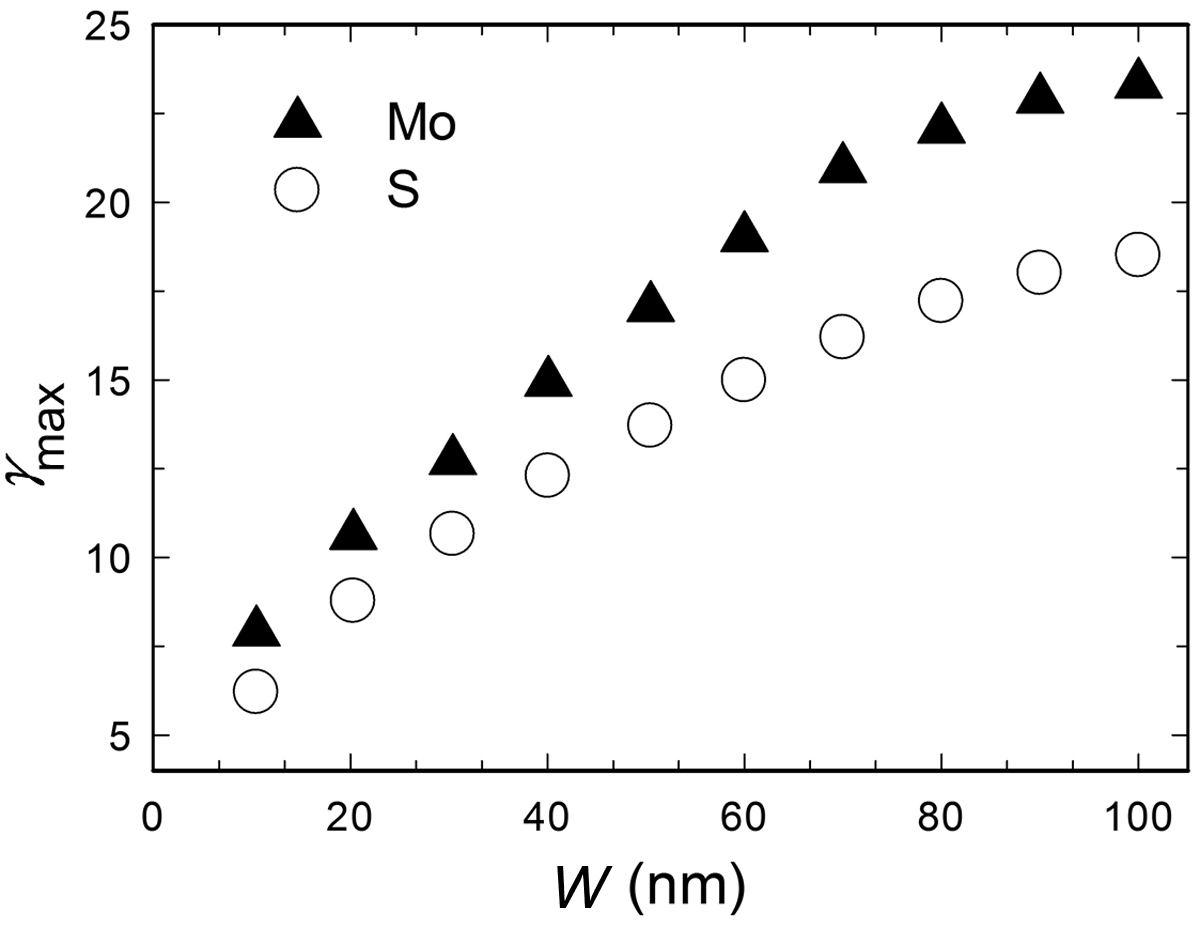}}
\caption{\label{F6}
Maximal charge enhancement ratio $\gamma_{max}$ versus $W$ for S and Mo atoms, respectively. $\gamma_{max}$ is defined as the ratio between $\eta$ at the ribbon edge and that at the ribbon center.
}
\end{figure}

The maximal charge enhancement ratio $\gamma_{max}$ is defined as the ratio of the maximal atomic charge density (at the edge) over the mean. The size-dependence of $\gamma_{max}$ for S and Mo is demonstrated in Fig.\ref{F6}. It can be seen that $\gamma_{max}$ increases with $W$ in decreasing proportionality. It can be seen on Fig.\ref{F6} that $\gamma_{max}$ is higher for Mo than for S. This could be due to the combination of the effect of the difference of electronegativities and the fact that there are roughly twice more S atoms than Mo atoms. Note, that a similar charge enhancement effects is also observable in spherical monolayer MoS$_{2}$ nano-flakes, as shown in supplementary material.

\section{Conclusions}
We predict charge enhancement effects in monolayer MoS$_{2}$ nanoribbons using an atomistic model, which is parametrized for predicting the distribution of doping electric charges. This model mimics each atom/ion as an induced dipole plus a quantity of Gaussian-distributed net charge. The equilibrium distributions of the net charges and induced dipoles are determined by minimizing the total electrostatic potential energy with the constrain of a fixed total electric charge. The parameters are obtained by empirically fitting to DFT calculation results. The charge distributions obtained by the charge-dipole model are compared with those obtained by another set of DFT calculations, by which good agreement is achieved. Different charge enhancement ratios are determined for S and Mo atoms, as a feature of MoS$_{2}$ distinct from graphene. The combination of this model with empirical force fields will enable large-scale atomistic simulations on electromechanical effects in layered MoS$_{2}$.

\section{SUPPLEMENTARY MATERIAL}
See supplementary material for calculation details about benchmarks, Gaussian characteristic width, electronegativity and charge enhancement in circular MoS$_{2}$, respectively.

\section{ACKNOWLEDGEMENTS}
Eric Duverger, Dan Huang and Gaoyang Gou are acknowledged for fruitful discussions. This work is supported by the Guangxi Key Laboratory Foundation (15-140-54), and Scientific Research Foundation of Guangxi University (XTZ160532).


\begin{thebibliography}{39}
\expandafter\ifx\csname natexlab\endcsname\relax\def\natexlab#1{#1}\fi
\expandafter\ifx\csname bibnamefont\endcsname\relax
  \def\bibnamefont#1{#1}\fi
\expandafter\ifx\csname bibfnamefont\endcsname\relax
  \def\bibfnamefont#1{#1}\fi
\expandafter\ifx\csname citenamefont\endcsname\relax
  \def\citenamefont#1{#1}\fi
\expandafter\ifx\csname url\endcsname\relax
  \def\url#1{\texttt{#1}}\fi
\expandafter\ifx\csname urlprefix\endcsname\relax\def\urlprefix{URL }\fi
\providecommand{\bibinfo}[2]{#2}
\providecommand{\eprint}[2][]{\url{#2}}

\bibitem[{\citenamefont{Butler et~al.}(2013)\citenamefont{Butler, Hollen, Cao, Cui,
  Gupta, Gutierrez, Heinz, Hong, Huang, Ismach, Johnston-Halperin, Kuno, Plashnitsa, Robinson, Ruoff, Salahuddin, Shan, Shi, Spencer, Terrones, Windl, and Goldberger}}]{Butler2013}
\bibinfo{author}{\bibfnamefont{S.~Z.}~\bibnamefont{Butler}},
\bibinfo{author}{\bibfnamefont{S.~M.}~\bibnamefont{Hollen}},
\bibinfo{author}{\bibfnamefont{L.~Y.}~\bibnamefont{Cao}},
\bibinfo{author}{\bibfnamefont{Y.}~\bibnamefont{Cui}},
\bibinfo{author}{\bibfnamefont{J.~A.}~\bibnamefont{Gupta}},
\bibinfo{author}{\bibfnamefont{H.~R.}~\bibnamefont{Guti\'errez}},
\bibinfo{author}{\bibfnamefont{T.~F.}~\bibnamefont{Heinz}},
\bibinfo{author}{\bibfnamefont{S.~S.}~\bibnamefont{Hong}},
\bibinfo{author}{\bibfnamefont{J.~X.}~\bibnamefont{Huang}},
\bibinfo{author}{\bibfnamefont{A.~F.}~\bibnamefont{Ismach}},
\bibinfo{author}{\bibfnamefont{E.~J.}~\bibnamefont{Halperin}},
\bibinfo{author}{\bibfnamefont{M.}~\bibnamefont{Kuno}},
\bibinfo{author}{\bibfnamefont{V.~V.}~\bibnamefont{Plashnitsa}},
\bibinfo{author}{\bibfnamefont{R.~D.}~\bibnamefont{Robinson}},
\bibinfo{author}{\bibfnamefont{R.~S.}~\bibnamefont{Rouff}},
\bibinfo{author}{\bibfnamefont{S.}~\bibnamefont{Salahuddin}},
\bibinfo{author}{\bibfnamefont{J.}~\bibnamefont{Shan}},
\bibinfo{author}{\bibfnamefont{L.}~\bibnamefont{Shi}},
\bibinfo{author}{\bibfnamefont{M.~G.}~\bibnamefont{Spencer}},
\bibinfo{author}{\bibfnamefont{M.}~\bibnamefont{Terrones}},
\bibinfo{author}{\bibfnamefont{W.}~\bibnamefont{Windl}}and
\bibinfo{author}{\bibfnamefont{J.~E.}~\bibnamefont{Goldberger}},
\bibinfo{journal}{ACS Nano} \textbf{\bibinfo{volume}{7}},
  \bibinfo{pages}{2898} (\bibinfo{year}{2013}).

  \bibitem[{\citenamefont{Huang et~al.}(2015)\citenamefont{Huang, Guo, Kang, Ai,
  and Li}}]{Huang2015}
\bibinfo{author}{\bibfnamefont{Y.}~\bibnamefont{Huang}},
  \bibinfo{author}{\bibfnamefont{J.}~\bibnamefont{Guo}},
  \bibinfo{author}{\bibfnamefont{Y.}~\bibnamefont{Kang}},
  \bibinfo{author}{\bibfnamefont{Y.}~\bibnamefont{Ai}}, \bibnamefont{and}
  \bibinfo{author}{\bibfnamefont{C.~M.} \bibnamefont{Li}},
  \bibinfo{journal}{Nanoscale} \textbf{\bibinfo{volume}{7}},
  \bibinfo{pages}{19358} (\bibinfo{year}{2015}).
  \bibitem[{\citenamefont{Acerce et~al.}(2017)\citenamefont{Acerce, Akdo?an, and
  Chhowalla}}]{Acerce2017}
\bibinfo{author}{\bibfnamefont{M.}~\bibnamefont{Acerce}},
  \bibinfo{author}{\bibfnamefont{E.}~\bibnamefont{Akdogan}}, \bibnamefont{and}
  \bibinfo{author}{\bibfnamefont{M.}~\bibnamefont{Chhowalla}},
  \bibinfo{journal}{Nature} \textbf{\bibinfo{volume}{549}},
  \bibinfo{pages}{370} (\bibinfo{year}{2017}).
 \bibitem[{\citenamefont{Lee et~al.}(2013)\citenamefont{Lee, Wang, He, Shan, and
  Feng}}]{Lee2013}
\bibinfo{author}{\bibfnamefont{J.}~\bibnamefont{Lee}},
  \bibinfo{author}{\bibfnamefont{Z.}~\bibnamefont{Wang}},
  \bibinfo{author}{\bibfnamefont{K.}~\bibnamefont{He}},
  \bibinfo{author}{\bibfnamefont{J.}~\bibnamefont{Shan}}, \bibnamefont{and}
  \bibinfo{author}{\bibfnamefont{P. X.-L.} \bibnamefont{Feng}},
  \bibinfo{journal}{ACS Nano} \textbf{\bibinfo{volume}{7}},
  \bibinfo{pages}{6086} (\bibinfo{year}{2013}).

\bibitem[{\citenamefont{Maity et~al.}(2017)\citenamefont{Maity, Mahanty, Sinha,
  Garain, Biswas, Ghosh, Manna, Ray, and Mandal}}]{Maity2017}
\bibinfo{author}{\bibfnamefont{K.}~\bibnamefont{Maity}},
  \bibinfo{author}{\bibfnamefont{B.}~\bibnamefont{Mahanty}},
  \bibinfo{author}{\bibfnamefont{T.}~\bibnamefont{Sinha}},
  \bibinfo{author}{\bibfnamefont{S.}~\bibnamefont{Garain}},
  \bibinfo{author}{\bibfnamefont{A.}~\bibnamefont{Biswas}},
  \bibinfo{author}{\bibfnamefont{S.}~\bibnamefont{Ghosh}},
  \bibinfo{author}{\bibfnamefont{S.}~\bibnamefont{Manna}},
  \bibinfo{author}{\bibfnamefont{S.}~\bibnamefont{Ray}}, \bibnamefont{and}
  \bibinfo{author}{\bibfnamefont{D.}~\bibnamefont{Mandal}},
  \bibinfo{journal}{Energy Tech.} \textbf{\bibinfo{volume}{5}},
  \bibinfo{pages}{234} (\bibinfo{year}{2017}).

\bibitem[{\citenamefont{Acerce et~al.}(2015)\citenamefont{Acerce, Voiry, and
  Chhowalla}}]{Acerce2015}
\bibinfo{author}{\bibfnamefont{M.}~\bibnamefont{Acerce}},
  \bibinfo{author}{\bibfnamefont{D.}~\bibnamefont{Voiry}}, \bibnamefont{and}
  \bibinfo{author}{\bibfnamefont{M.}~\bibnamefont{Chhowalla}},
  \bibinfo{journal}{Nature Nanotech.} \textbf{\bibinfo{volume}{10}},
  \bibinfo{pages}{313} (\bibinfo{year}{2015}).

  \bibitem[{\citenamefont{Wu et~al.}(2009)\citenamefont{Wu, Pei, Ren, Tang, Gao,
  Liu, Li, Liu, and Cheng}}]{Wu2009a}
\bibinfo{author}{\bibfnamefont{Z.}~\bibnamefont{Wu}},
  \bibinfo{author}{\bibfnamefont{S.}~\bibnamefont{Pei}},
  \bibinfo{author}{\bibfnamefont{W.}~\bibnamefont{Ren}},
  \bibinfo{author}{\bibfnamefont{D.}~\bibnamefont{Tang}},
  \bibinfo{author}{\bibfnamefont{L.}~\bibnamefont{Gao}},
  \bibinfo{author}{\bibfnamefont{B.}~\bibnamefont{Liu}},
  \bibinfo{author}{\bibfnamefont{F.}~\bibnamefont{Li}},
  \bibinfo{author}{\bibfnamefont{C.}~\bibnamefont{Liu}}, \bibnamefont{and}
  \bibinfo{author}{\bibfnamefont{H.}~\bibnamefont{Cheng}},
  \bibinfo{journal}{Adv. Mater.} \textbf{\bibinfo{volume}{21}},
  \bibinfo{pages}{1756} (\bibinfo{year}{2009}).

  \bibitem[{\citenamefont{Gartstein et~al.}(2003)\citenamefont{Gartstein,
  Zakhidov, and Baughman}}]{Gartstein2003}
\bibinfo{author}{\bibfnamefont{Y.~N.} \bibnamefont{Gartstein}},
  \bibinfo{author}{\bibfnamefont{A.~A.} \bibnamefont{Zakhidov}},
  \bibnamefont{and} \bibinfo{author}{\bibfnamefont{R.~H.}
  \bibnamefont{Baughman}}, \bibinfo{journal}{Phys. Rev. B}
  \textbf{\bibinfo{volume}{68}}, \bibinfo{pages}{115415}
  (\bibinfo{year}{2003}).

\bibitem[{\citenamefont{Jung et~al.}(2009)\citenamefont{Jung, Kim, Jockusch, Turro,
  Kim, and Brus}}]{Jung2009}
\bibinfo{author}{\bibfnamefont{N.}~\bibnamefont{Jung}},
  \bibinfo{author}{\bibfnamefont{N.}~\bibnamefont{Kim}},
  \bibinfo{author}{\bibfnamefont{S.}~\bibnamefont{Jockusch}},
  \bibinfo{author}{\bibfnamefont{N.~J.}~\bibnamefont{Turro}},
   \bibinfo{author}{\bibfnamefont{P.}~\bibnamefont{Kim}},
  \bibnamefont{and}
  \bibinfo{author}{\bibfnamefont{L.} \bibnamefont{Brus}},
  \bibinfo{journal}{Nano Lett.} \textbf{\bibinfo{volume}{9}},
  \bibinfo{pages}{4133} (\bibinfo{year}{2009}).

    \bibitem[{\citenamefont{Guinea}(2007)\citenamefont{Guinea}}]{Guinea2007}
\bibinfo{author}{\bibfnamefont{F.}~\bibnamefont{Guinea}},
  \bibinfo{journal}{Phys. Rev. B} \textbf{\bibinfo{volume}{75}},
  \bibinfo{pages}{235433} (\bibinfo{year}{2007}).
	
 \bibitem[{\citenamefont{Purcell et~al.}(2002)\citenamefont{Purcell et~al.}}]{Purcell2002}
\bibinfo{author}{\bibfnamefont{S. T.}~\bibnamefont{Purcell}},
  \bibinfo{author}{\bibfnamefont{P.}~\bibnamefont{Vincent}},
  \bibinfo{author}{\bibfnamefont{C.}~\bibnamefont{Journet}},
	  \bibnamefont{and}
  \bibinfo{author}{\bibfnamefont{V. T.}~\bibnamefont{Binh}},
  \bibinfo{journal}{Phys. Rev. Lett.} \textbf{\bibinfo{volume}{88}},
  \bibinfo{pages}{105502} (\bibinfo{year}{2002}).
	
   \bibitem[{\citenamefont{Brunel et~al.}(2010)\citenamefont{Brunel, Mayer, and Melin}}]{Brunel2010}
\bibinfo{author}{\bibfnamefont{D.}~\bibnamefont{Brunel}},
\bibinfo{author}{\bibfnamefont{A.}~\bibnamefont{Mayer}}, \bibnamefont{and}
  \bibinfo{author}{\bibfnamefont{T.}~\bibnamefont{M\'{e}lin}},
  \bibinfo{journal}{ACS Nano} \textbf{\bibinfo{volume}{4}},
  \bibinfo{pages}{5978} (\bibinfo{year}{2010}).
	
    \bibitem[{\citenamefont{Datta et~al.}({2009})\citenamefont{Datta, Strachan, Mele,
  and Johnson}}]{Datta2009}
\bibinfo{author}{\bibfnamefont{S.~S.}~\bibnamefont{Datta}},
  \bibinfo{author}{\bibfnamefont{D.~R.}~\bibnamefont{Strachan}},
  \bibinfo{author}{\bibfnamefont{E.~J.}~\bibnamefont{Mele}}, \bibnamefont{and}
  \bibinfo{author}{\bibfnamefont{A.~T.~C.}~\bibnamefont{Johnson}},
  \bibinfo{journal}{Nano Lett.} \textbf{\bibinfo{volume}{{9}}},
  \bibinfo{pages}{7} (\bibinfo{year}{{2008}}).

  \bibitem[{\citenamefont{Wang et~al.}(2008)\citenamefont{Wang, Zdrojek,
  M\'{e}lin, and Devel}}]{Wang2008c}
\bibinfo{author}{\bibfnamefont{Z.}~\bibnamefont{Wang}},
  \bibinfo{author}{\bibfnamefont{M.}~\bibnamefont{Zdrojek}},
  \bibinfo{author}{\bibfnamefont{T.}~\bibnamefont{M\'{e}lin}},
  \bibnamefont{and} \bibinfo{author}{\bibfnamefont{M.}~\bibnamefont{Devel}},
  \bibinfo{journal}{Phys. Rev. B} \textbf{\bibinfo{volume}{78}},
  \bibinfo{pages}{085425} (\bibinfo{year}{2008}).
  
  
  
  
  
  
  
  
    \bibitem[{\citenamefont{Wang }(2009)\citenamefont{Wang, }}]{Wang2009}
\bibinfo{author}{\bibfnamefont{Z.}~\bibnamefont{Wang}},
  \bibinfo{journal}{Phys. Rev. B} \textbf{\bibinfo{volume}{79}},
  \bibinfo{pages}{155407} (\bibinfo{year}{2009}).
  
  
  
  
  
  
  
  
  
  

  \bibitem[{\citenamefont{Keblinski et~al.}(2002)\citenamefont{Keblinski, Nayak,
  Zapol, and Ajayan}}]{Keblinski2002}
\bibinfo{author}{\bibfnamefont{P.}~\bibnamefont{Keblinski}},
  \bibinfo{author}{\bibfnamefont{S.~K.} \bibnamefont{Nayak}},
  \bibinfo{author}{\bibfnamefont{P.}~\bibnamefont{Zapol}}, \bibnamefont{and}
  \bibinfo{author}{\bibfnamefont{P.~M.} \bibnamefont{Ajayan}},
  \bibinfo{journal}{Phys. Rev. Lett.} \textbf{\bibinfo{volume}{89}},
  \bibinfo{pages}{255503} (\bibinfo{year}{2002}).
	
	  \bibitem[{\citenamefont{Brennan et al.}(2017)\citenamefont{Brennan}}]{Brennan2017}
\bibinfo{author}{\bibfnamefont{C.~J.}~\bibnamefont{Brennan}},
  \bibinfo{author}{\bibfnamefont{R.} \bibnamefont{Ghosh}},
  \bibinfo{author}{\bibfnamefont{K.}~\bibnamefont{Koul}},
	\bibinfo{author}{\bibfnamefont{S.~K.}~\bibnamefont{Banerjee}},
	\bibinfo{author}{\bibfnamefont{N.}~\bibnamefont{Lu}},
	\bibnamefont{and}
  \bibinfo{author}{\bibfnamefont{E.~T.} \bibnamefont{Yu}},
  \bibinfo{journal}{Nano Lett.} \textbf{\bibinfo{volume}{17}},
  \bibinfo{pages}{5464} (\bibinfo{year}{2017}).
		
\bibitem[{\citenamefont{Hartman et~al.}(2004)\citenamefont{Hartman, Jouzi,
  Barnett, and Xu}}]{Hartman2004}
\bibinfo{author}{\bibfnamefont{A.~Z.}~\bibnamefont{Hartman}},
  \bibinfo{author}{\bibfnamefont{M.}~\bibnamefont{Jouzi}},
  \bibinfo{author}{\bibfnamefont{R.~L.}~\bibnamefont{Barnett}}, \bibnamefont{and}
  \bibinfo{author}{\bibfnamefont{J.~M.}~\bibnamefont{Xu}}, \bibinfo{journal}{Phys.
  Rev. Lett.} \textbf{\bibinfo{volume}{92}}, \bibinfo{pages}{236804}
  (\bibinfo{year}{2004}).
  
  
  
  

  
  
  \bibitem[{\citenamefont{Wang et~al.}(2009)\citenamefont{wang
  }}]{Wange2009}
\bibinfo{author}{\bibfnamefont{Z.}~\bibnamefont{Wang}},
\bibinfo{journal}{Carbon}
   \textbf{\bibinfo{volume}{47}}, \bibinfo{pages}{3050}
  (\bibinfo{year}{2009}).
  
  
  
  
  
  
  
  
  \bibitem[{\citenamefont{wang et~al.}(2004)\citenamefont{Wang, Philippe
 }}]{Wangd2009}
\bibinfo{author}{\bibfnamefont{Z.}~\bibnamefont{Wang}},\bibnamefont{and}
  \bibinfo{author}{\bibfnamefont{L.}~\bibnamefont{Philippe}}, \bibinfo{journal}{Phys.
  Rev. Lett.} \textbf{\bibinfo{volume}{102}}, \bibinfo{pages}{215501}
  (\bibinfo{year}{2009}).
  
  
  
  
  
  
  
  
	
\bibitem[{\citenamefont{Bennett et~al.}(2010)\citenamefont{Bennett, Cockins,
  Miyahara, Grutter, and Clerk}}]{Bennett2010}
\bibinfo{author}{\bibfnamefont{S.~D.} \bibnamefont{Bennett}},
  \bibinfo{author}{\bibfnamefont{L.}~\bibnamefont{Cockins}},
  \bibinfo{author}{\bibfnamefont{Y.}~\bibnamefont{Miyahara}},
  \bibinfo{author}{\bibfnamefont{P.}~\bibnamefont{Grutter}}, \bibnamefont{and}
  \bibinfo{author}{\bibfnamefont{A.~A.} \bibnamefont{Clerk}},
  \bibinfo{journal}{Phys. Rev. Lett.} \textbf{\bibinfo{volume}{104}},
  \bibinfo{pages}{017203} (\bibinfo{year}{2010}).

\bibitem[{\citenamefont{Wang et~al.}({2007})\citenamefont{Wang, Devel, Langlet,
  and Dulmet}}]{Wang2007}
\bibinfo{author}{\bibfnamefont{Z.}~\bibnamefont{Wang}},
  \bibinfo{author}{\bibfnamefont{M.}~\bibnamefont{Devel}},
  \bibinfo{author}{\bibfnamefont{R.}~\bibnamefont{Langlet}}, \bibnamefont{and}
  \bibinfo{author}{\bibfnamefont{B.}~\bibnamefont{Dulmet}},
  \bibinfo{journal}{Phys. Rev. B} \textbf{\bibinfo{volume}{{75}}},
   \bibinfo{pages}{205414} (\bibinfo{year}{{2007}}).

\bibitem[{\citenamefont{Wang and Devel}({2007})}]{Wang2007a}
\bibinfo{author}{\bibfnamefont{Z.}~\bibnamefont{Wang}} \bibnamefont{and}
  \bibinfo{author}{\bibfnamefont{M.}~\bibnamefont{Devel}},
  \bibinfo{journal}{Phys. Rev. B} \textbf{\bibinfo{volume}{{76}}},
    \bibinfo{pages}{195434} (\bibinfo{year}{{2007}}).

\bibitem[{\citenamefont{Applequist et~al.}(1972)\citenamefont{Applequist, Carl,
  and Fung}}]{Applequist1972}
\bibinfo{author}{\bibfnamefont{J.}~\bibnamefont{Applequist}},
  \bibinfo{author}{\bibfnamefont{J.}~\bibnamefont{Carl}}, \bibnamefont{and}
  \bibinfo{author}{\bibfnamefont{K.}~\bibnamefont{Fung}}, \bibinfo{journal}{J.
  Am. Chem. Soc.} \textbf{\bibinfo{volume}{94}}, \bibinfo{pages}{2952}
  (\bibinfo{year}{1972}).
	
\bibitem[{\citenamefont{Jensen et~al.}(2002)\citenamefont{L. Jensen, P.-O. \AA{}strand, A. Osted, J. Kongsted, and K. V. Mikkelsen}}]{Jensen2002}
\bibinfo{author}{\bibfnamefont{L.}~\bibnamefont{Jensen}},
  \bibinfo{author}{\bibfnamefont{P.-O.}~\bibnamefont{\AA{}strand}},
  \bibinfo{author}{\bibfnamefont{A.}~\bibnamefont{Osted}},
	\bibinfo{author}{\bibfnamefont{J.}~\bibnamefont{Kongsted}}, \bibnamefont{and}
  \bibinfo{author}{\bibfnamefont{K. V.}~\bibnamefont{Mikkelsen}},
  \bibinfo{journal}{J. Chem. Phys.} \textbf{\bibinfo{volume}{116}},
  \bibinfo{pages}{4001} (\bibinfo{year}{2002}).

\bibitem[{\citenamefont{Jensen et~al.}(2004)\citenamefont{Jensen, Astrand, and
  Mikkelsen}}]{jensen2004}
\bibinfo{author}{\bibfnamefont{L.}~\bibnamefont{Jensen}},
  \bibinfo{author}{\bibfnamefont{P.}~\bibnamefont{\AA{}strand}}, \bibnamefont{and}
  \bibinfo{author}{\bibfnamefont{K.}~\bibnamefont{Mikkelsen}},
  \bibinfo{journal}{J. Phys. Chem. A} \textbf{\bibinfo{volume}{108}},
  \bibinfo{pages}{8795} (\bibinfo{year}{2004}).

  \bibitem[{\citenamefont{Mayer et~al.}(2006)\citenamefont{Mayer, Lambin, and Langlet}}]{Mayer2006}
\bibinfo{author}{\bibfnamefont{A.}~\bibnamefont{Mayer}},
  \bibinfo{author}{\bibfnamefont{P.}~\bibnamefont{Lambin}}, \bibnamefont{and}
  \bibinfo{author}{\bibfnamefont{R.}~\bibnamefont{Langlet}},
  \bibinfo{journal}{Appl. Phys. Lett.} \textbf{\bibinfo{volume}{89}},
  \bibinfo{pages}{063117} (\bibinfo{year}{2006}).
	
\bibitem[{\citenamefont{Kresse and Furthmuller}(1996)}]{Kresse1996}
\bibinfo{author}{\bibfnamefont{G.}~\bibnamefont{Kresse}} \bibnamefont{and}
  \bibinfo{author}{\bibfnamefont{J.}~\bibnamefont{Furthmuller}},
  \bibinfo{journal}{Phys. Rev. B} \textbf{\bibinfo{volume}{54}},
  \bibinfo{pages}{11169} (\bibinfo{year}{1996}).

\bibitem[{\citenamefont{Kresse and Joubert}(1999)}]{Kresse1999}
\bibinfo{author}{\bibfnamefont{G.}~\bibnamefont{Kresse}} \bibnamefont{and}
  \bibinfo{author}{\bibfnamefont{D.}~\bibnamefont{Joubert}},
  \bibinfo{journal}{Phys. Rev. B} \textbf{\bibinfo{volume}{59}},
  \bibinfo{pages}{1758} (\bibinfo{year}{1999}).

\bibitem[{\citenamefont{Langlet et~al.}(2006)\citenamefont{Langlet, Devel, and
  Lambin}}]{Langlet2006}
\bibinfo{author}{\bibfnamefont{R.}~\bibnamefont{Langlet}},
  \bibinfo{author}{\bibfnamefont{M.}~\bibnamefont{Devel}}, \bibnamefont{and}
  \bibinfo{author}{\bibfnamefont{P.}~\bibnamefont{Lambin}},
  \bibinfo{journal}{Carbon} \textbf{\bibinfo{volume}{44}},
  \bibinfo{pages}{2883} (\bibinfo{year}{2006}).	

 \bibitem[{\citenamefont{A. Mayer and P.-O. Astrand }(2008)}]{Mayer2008}
  \bibinfo{author}{\bibfnamefont{A.}~\bibnamefont{Mayer}} \bibnamefont{and}
  \bibinfo{author}{\bibfnamefont{P.-O.}~\bibnamefont{\AA{}strand}},
  \bibinfo{journal}{J. Phys. Chem. A} \textbf{\bibinfo{volume}{112}},
  \bibinfo{pages}{1277} (\bibinfo{year}{2008}).	 	
    
\bibitem[{\citenamefont{J. Chen and J. Mart\'inez}(2007)}]{Chen2007}
  \bibinfo{author}{\bibfnamefont{J.}~\bibnamefont{Chen}} \bibnamefont{and}
  \bibinfo{author}{\bibfnamefont{J.}~\bibnamefont{Mart\'inez}},
  \bibinfo{journal}{Chem. Phys. Lett.} \textbf{\bibinfo{volume}{438}},
  \bibinfo{pages}{315} (\bibinfo{year}{2007}).	 

\bibitem[{\citenamefont{J.Chen et~al.}(2008)\citenamefont{Jiahao, Hundertmark, and
  Mart\'inez}}]{Chen2008}
\bibinfo{author}{\bibfnamefont{J.}~\bibnamefont{Chen}},
  \bibinfo{author}{\bibfnamefont{D.}~\bibnamefont{Hundertmark}}, \bibnamefont{and}
  \bibinfo{author}{\bibfnamefont{T.}~\bibnamefont{Mart\'inez}},
  \bibinfo{journal}{J. Chem. Phys.} \textbf{\bibinfo{volume}{129}},
  \bibinfo{pages}{214113} (\bibinfo{year}{2008}).	           
 
   \bibitem[{\citenamefont{J.Bredas et~al.}(2004)\citenamefont{Bredas, Beljonne, Coropceanu, and
Cornil  }}]{Bredas2004}
\bibinfo{author}{\bibfnamefont{J.}~\bibnamefont{Br\'edas}},
  \bibinfo{author}{\bibfnamefont{D.}~\bibnamefont{Beljonne}},
  \bibinfo{author}{\bibfnamefont{V.}~\bibnamefont{Coropceanu}}, \bibnamefont{and}
    \bibinfo{author}{\bibfnamefont{J.}~\bibnamefont{Cornil}},
  \bibinfo{journal}{Chem. Rev.} \textbf{\bibinfo{volume}{104}},
  \bibinfo{pages}{4971} (\bibinfo{year}{2004}).	
  
 \bibitem[{\citenamefont{S. Kummel}(2004)\citenamefont{Kummel }}]{Kummel2017}
\bibinfo{author}{\bibfnamefont{S.}~\bibnamefont{K\"ummel}},
  \bibinfo{journal}{Adv. Energy Mater.} \textbf{\bibinfo{volume}{7}},
  \bibinfo{pages}{1700440} (\bibinfo{year}{2017}).	   
  
\bibitem[{\citenamefont{Mayer}(2007)}]{Mayer2007}
\bibinfo{author}{\bibfnamefont{A.}~\bibnamefont{Mayer}},
  \bibinfo{journal}{Phys. Rev. B} \textbf{\bibinfo{volume}{75}},
  \bibinfo{pages}{045407} (\bibinfo{year}{2007}).

  \bibitem[{\citenamefont{Ma}(2006)}]{Ma2006}
\bibinfo{author}{\bibfnamefont{Y.}~\bibnamefont{Ma}} \bibnamefont{and}
\bibinfo{author}{\bibfnamefont{S.H.}~\bibnamefont{Garofalini}},
  \bibinfo{journal}{J. Chem. Phys.} \textbf{\bibinfo{volume}{124}},
  \bibinfo{pages}{234102} (\bibinfo{year}{2006}).
\bibitem[{\citenamefont{Henkelman et~at}(2006)}]{Henkelman2006}
\bibinfo{author}{\bibfnamefont{G.}~\bibnamefont{Henkelman}}, \bibnamefont
  \bibinfo{author}{\bibfnamefont{A.}~\bibnamefont{Arnaldsson}}, {and}
    \bibinfo{author}{\bibfnamefont{H.}~\bibnamefont{J\'onsson}},
  \bibinfo{journal}{Comp. Mater. Sci.} \textbf{\bibinfo{volume}{36}},
  \bibinfo{pages}{354} (\bibinfo{year}{2006}).
  \bibitem[{\citenamefont{Wang and Scharstein}(2010)}]{zw2010d}
\bibinfo{author}{\bibfnamefont{Z.}~\bibnamefont{Wang}} \bibnamefont{and}
  \bibinfo{author}{\bibfnamefont{R.~W.} \bibnamefont{Scharstein}},
  \bibinfo{journal}{Chem. Phy. Lett.} \textbf{\bibinfo{volume}{489}},
  \bibinfo{pages}{229} (\bibinfo{year}{2010}).
	
  \end{thebibliography}
\end{document}